\documentclass[intlimits,twoside,a4paper]{article}

\usepackage{amsmath,amssymb}
\usepackage{graphicx}

\usepackage[T2A]{fontenc}
\usepackage[cp1251]{inputenc}

\usepackage{dcolumn}

\usepackage{cmpj2}

\issue{2015}{18}{4}{43701}
\doinumber{10.5488/CMP.18.43701}

\title[Cu$^{2+}$ center  in Cd$_{2}$(NH$_{4}$)$_{2}$(SO$_{4}$)$_{3}$ single crystal]{Theoretical studies of the local structures  and EPR parameters for Cu$^{2+}$ center  in Cd$_{2}$(NH$_{4}$)$_{2}$(SO$_{4}$)$_{3}$ \\ single crystal}

\author[Ch.-Y. Li \textsl{et al.}]{Ch.-Y. Li\thanks{E-mail: cyli1962@126.com}\,, L.-B. Chen, J.-J. Mao, X.-M. Zheng}
\address{School of Physics and Electronic Information, Shangrao
Normal University, Shangrao Jiangxi 334000, P. R. China}

\date{Received March 24, 2015, in final form July 10, 2015}
\authorcopyright{Ch.-Y. Li, L.-B. Chen, J.-J. Mao, X.-M. Zheng, 2015}

\begin{document}

\maketitle

\begin{abstract}

The electron paramagnetic resonance (EPR) parameters
($g$ factors $g_{i}$ and the hyperfine
structure constants ${{A}}_{{i}}$, ${i} =
{x}, {y}, {z}$) are interpreted by using the
perturbation formulae for a $3{d}^{9}$ ion in rhombically
({D}$_\textrm{2h}$) elongated octahedra. In the calculated formulae,
the crystal field parameters are set up from the superposition model,
and the contribution to the EPR parameters from the
admixture of $d$-orbitals in the ground state wave function of the
Cu$^{2+}$ ion was taken into account. Based on the
calculation, local structural parameters of the impurity
Cu$^{2+}$ center in
Cd$_{2}$(NH$_{4}$)$_{2}$(SO$_{4}$)$_{3}$ (CAS)
crystal were obtained (i.e., ${R}_{{x}}\approx 2.05$~{\AA}, ${R}_{{y}} \approx 1.91$~{\AA},
${R}_{{z}} \approx 2.32$~{\AA}). The theoretical
EPR parameters based on the above
Cu$^{2+}$--$\text{O}^{2-}$ bond lengths in CAS
crystal show a good agreement with the observed values.
The results are discussed.
 \\ Cu$^{2+}$ doping
\pacs 76.30.Fc, 75.10.Dg, 71.70.ch
\end{abstract}

\section{Introduction}

Single crystal
Cd$_{2}$(NH$_{4}$)$_{2}$(SO$_{4}$)$_{3}$ (CAS)
has attracted interest of researchers due to the unique dielectric
\cite{1}, phase transition \cite{2}, optical \cite{3}, birefringent
and electrooptical properties \cite{4}. The above properties may be
closely related to the local structures, chemical bonding and
electronic states of the doped ions in the hosts. Electron
paramagnetic resonance (EPR) has long been considered as
one of the most useful tools for the experimental study of chemical
bonding. The EPR method provides a detailed description of
the ground state of paramagnetic ions and enables one to explain the
nature of crystal field and its symmetry produced by ligands around
the metal ion \cite{5,6,7}. Among these transition metal ions,
Cu$^{2+}$ ions with $3d^{9}$ configuration are widely
used as paramagnetic probes as they represent a relatively simple
one-hole magnetic system which can be used to obtain information
regarding the electron wave function when there is a ligand field of
low symmetry. Thus, EPR spectra of Cu$^{2+}$ ion in
different diamagnetic host lattices have been studied by many
workers to get some data on the structure, dynamics and
environment of host lattices \cite{8,9,10,11,12}. For example, the
EPR experiments were carried out for Cu$^{2+}$ doped
in CAS single crystal, and the EPR parameters
(anisotropic $g$ factors $g_{i}$ and the
hyperfine structure constants ${{A}}_{{i}}$,
${i} = {x}, {y}, {z}$) were also measured for rhombic
Cu$^{2+}$ center \cite{13}. However, no
satisfactory interpretation to the above experimental results has
been made so far, and the data on the defect structures of
Cu$^{2+}$ center have not been obtained yet.

Considering that ({i}) the data on local
structures and electronic states for Cu$^{2+}$ in the
CAS single crystal would be helpful in understanding the
microscopic mechanisms of EPR behaviors of this material
containing Cu$^{2+}$ dopants and ({ii}) the
anisotropic $g$ factors for a $d^{9}$ ion in crystals
are sensitive to its immediate environment (and hence to defect
structure of $d^{9}$ impurity center). Thus, further
investigations on EPR parameters and the defect structures
for the Cu$^{2+}$ center are of fundamental and practical
significance. In this paper, we have carried out local structure
calculations for a paramagnetic Cu$^{2+}$ center in
CAS and have interpreted the EPR parameters in
this system. The theoretical results are in good agreement with the
experimental values. The Cu$^{2+}$--$\text{O}^{2-}$ bond
lengths are obtained as follows: ${R}_{{x}} \approx  2.05$~{\AA}, ${R}_{{y}} \approx 1.91$~{\AA},
${R}_{{z}}\approx 2.32$~{\AA}.

\section{Calculation}

In the lattice of CAS crystal, each {Cd}$^{2+}$
is surrounded by six oxygen atoms which form a slightly distorted
octahedron \cite{13}. When Cu$^{2+}$ is doped in CAS
crystal, it enters the lattice at Cd$^{2+}$ site. For a
$3d^{9}$ (Cu$^{2+}$) ion in an octahedral complex with
a rhombic elongation it would give ${{g}}_{{z}} >
{{g}}_{{x}}$, ${{g}}_{{y}}> 2$
 \cite{12,14,15}. Experimental results of EPR parameters in
reference \cite{9} agree with this relation. That is to say, the studied
Cu(H$_{2}$O)$_{6}^{2+}$ clusters in CAS
crystal are in rhombically elongated octahedra. However, the host
Cd(H$_{2}$O)$_{6}^{2+}$ clusters in CAS
crystal are under rhombically compressed octahedra similar to many
other tutton \cite{16}. A Jahn-Teller
ion is due to Cu$^{2+}$. When it occupies a cubic or trigonal octahedral site, the
ground state is doublet $^{2}{E}$. The degeneracy of
$^{2}{E}$ state should be removed by Jahn-Teller effect, which
makes these octahedral {CuL}$_{6}$ clusters become rhombic.
Thus, the change of compressed
$\text{Cd}(\text{H}_{2}\text{O})_{6}^{2+}$ octahedra in the host
crystals to elongated $\text{Cu}(\text{H}_{2}\text{O})_{6}^{2+}$
octahedra in the impurity centers due to Jahn-Teller effect for
Cu$^{2+}$ doped CAS crystal becomes understandable.

The local structures of the impurity Cu$^{2+}$ center
in CAS single crystal under rhombic symmetry are described
by the metal-ligand distance ${{R}}_{{i}}$ (${i}
= {x}, {y}, {z}$). Then, from the superposition model
\cite{17} and local geometrical relationship of the impurity
Cu$^{2+}$ center, the rhombic field parameters ${D}_{s}$,
${D}_{t}$, ${D}_{\xi}$ and ${D}_{\eta}$ can be expressed
as follows:
\begin{eqnarray}
\label{eq1}
D_{s} &=&
\frac{4}{7}\overline{A}_{2}({R}_{0})\left[\left(\frac{{R}_{0}}{{R}_{x}}\right)^{t_{2}}+ \left(\frac{{R}_{0}}{{R}_{y}}\right)^{t_{2}}-2\left(\frac{{R}_{0}}{{R}_{z}}\right)^{t_{2}}\right],\nonumber\\
D_{t} &=&
\frac{8}{21}\overline{A}_{4}({R}_{0})\left[\left(\frac{{R}_{0}}{{R}_{x}}\right)^{t_{4}}+
\left(\frac{{R}_{0}}{{R}_{y}}\right)^{t_{4}}-2\left(\frac{{R}_{0}}{{R}_{z}}\right)^{t_{4}}\right],\nonumber\\
D_{\xi} &=&
\frac{2}{7}\overline{A}_{2}({R}_{0})\left[\left(\frac{{R}_{0}}{{R}_{x}}\right)^{t_{2}}-
\left(\frac{{R}_{0}}{{R}_{y}}\right)^{t_{2}}\right],\nonumber\\
D_{\eta} &=&
\frac{10}{21}\overline{A}_{4}({R}_{0})\left[\left(\frac{{R}_{0}}{{R}_{x}}\right)^{t_{4}}-
\left(\frac{{R}_{0}}{{R}_{y}}\right)^{t_{4}}\right].
\end{eqnarray}

Here, $t_{2}\approx 3$ and $t_{4}\approx 5$ are the power-law
exponents due to the dominant ionic nature of the bonds
\cite{9,12,18,19,20}. $\overline{A}_{2}({R}_{0})$, and
$\overline{A}_{4}({R}_{0})$ are intrinsic parameters with
the reference distance ${{R}}_{0}$ [taken as
${{R}}_{0} = \overline{{R}} =
({{R}}_{x}+{{R}}_{y}+{{R}}_{z})/3$].
The ratio
$\overline{A}_{2}({R}_{0}) / \overline{A}_{4}({R}_{0})$
was found to be in the range $9 - 12$ as per several studies
of optical and EPR spectra using superposition model for
$3d^{n}$ ions in crystals \cite{12,19,20,21,22}. Here, we
take
$\overline{A}_{2}({R}_{0})/\overline{A}_{4}({R}_{0})= 12$. Therefore, axial and perpendicular anisotropies
$\Delta g$ [$= {{g}}_{{z}} - ({{g}}_{{x}} + {{g}}_{{y}})/2$] and
$\delta g$ [$= {{g}}_{{y}} - {{g}}_{{x}}$] of EPR ${{g}}$
factors are correlated to the rhombic field parameters and hence to
the local structures of the  systems studied.

For a $3{d}^{9}$ (Cu$^{2+}$) ion under rhombically
elongated octahedra, its lower orbital doublet $^{2}{E}_{g}$
would be separated into two singlets $^{2}{A}_{1g}(\theta)$
and $^{2}{A}_{1g}(\epsilon)$. Meanwhile, the higher cubic
orbital triplet $^{2}{T}_{2g}$ would be split into three
singlets $^{2}{B}_{1g}(\zeta)$, $^{2}{B}_{2g}(\eta)$
and $^{2}{B}_{3g}(\xi)$ \cite{9}. Since the states
$^{2}{A}_{1g}(\theta)$ and $^{2}{A}_{1g}(\epsilon)$
belong to the same representation of rhombic symmetry group, the
ground state will be neither $^{2}{A}_{1g}(\theta)$ nor
$^{2}{A}_{1g}(\epsilon)$ but an admixture of both
\cite{8,22}, i.e.,
\begin{eqnarray}
\Phi= {N}\left[\alpha|d_{x^{2}-y^{2}}+\beta|d_{3z^{2}-r^{2}}\right],
\end{eqnarray}
where ${N}$ is the probability of finding electron in the metal
${d}$-orbital, the characteristic of covalency of the system.
$\alpha$ and $\beta$ are the mixing coefficients due to the rhombic
field components and satisfy the normalization relation:
\begin{eqnarray}
\alpha^{2}+\beta^{2} = 1.
\end{eqnarray}

From perturbation theory, the high-order perturbation formulae
of EPR parameters ($g$ factors
${{g}}_{{x}}$, ${{g}}_{{y}}$,
${{g}}_{{z}}$ and hyperfine structure constants
${{A}}_{{x}}$, ${{A}}_{{y}}$,
${{A}}_{{z}}$) for $3d^{9}$ ions in rhombic
symmetry can be expressed as~\cite{14,22}:
\begin{eqnarray}
g_{{x}}&=& g_{{s}} + \frac{2{k}\zeta\left(\alpha+\sqrt{3}\beta\right)}{E_{4}} - \frac{2\alpha{k}\zeta^{2}\left(\alpha+\sqrt{3}\beta\right)}{E_{2}E_{4}}
 + \frac{{k}\zeta^{2}\left(\alpha^{2}-3\beta^{2}\right)}{E_{3}E_{4}} \nonumber\\
&& - \frac{2\alpha^{2}g_{s}\zeta^{2}}{E_{2}^{2}}
 - \frac{g_{s}\zeta^{2}\left(\alpha-\sqrt{3}\beta\right)^{2}}{2E_{3}^{2}}
 + \frac{2\alpha{k}\zeta^{2}\left(\alpha-\sqrt{3}\beta\right)^{2}}{E_{2}E_{3}}\,,\nonumber\\
g_{{y}}&=& g_{{s}} +
\frac{2{k}\zeta\left(\alpha-\sqrt{3}\beta\right)}{E_{3}} -
\frac{2\alpha{k}\zeta^{2}\left(\alpha-\sqrt{3}\beta\right)}{E_{2}E_{3}}
 + \frac{{k}\zeta^{2}\left(\alpha^{2}-3\beta^{2}\right)}{E_{3}E_{4}} \nonumber\\
&& - \frac{2\alpha^{2}g_{s}\zeta^{2}}{E_{2}^{2}}
 - \frac{g_{s}\zeta^{2}\left(\alpha+\sqrt{3}\beta\right)^{2}}{2E_{4}^{2}}
 + \frac{2\alpha{k}\zeta^{2}\left(\alpha+\sqrt{3}\beta\right)^{2}}{E_{2}E_{4}}\,,\\
g_{{z}}&=& g_{{s}} +
\frac{8\alpha^{2}{k}\zeta}{E_{2}} -
\frac{2\alpha{k}\zeta^{2}\left(\alpha-\sqrt{3}\beta\right)}{E_{2}E_{3}} -
\frac{2\alpha{k}\zeta^{2}\left(\alpha+\sqrt{3}\beta\right)}{E_{2}E_{4}}\nonumber\\
&& - \frac{g_{s}\zeta^{2}\left(\alpha-\sqrt{3}\beta\right)}{2E_{3}^{2}} - \frac{g_{s}\zeta^{2}\left(\alpha+\sqrt{3}\beta\right)^{2}}{2E_{4}^{2}}
 - \frac{{k}\zeta^{2}\left(\alpha-3\beta^{2}\right)}{E_{3}E_{4}},\nonumber
\end{eqnarray}
\begin{eqnarray}
A_{x}&=&P_{0}\left[-\kappa + \frac{2N^{2}}{7} - \kappa^{\prime} + \left(g_{x} - g_{s}\right) - \frac{3}{14} \left(g_{y} - g_{s}\right)\right],\nonumber\\
A_{y}&=&P_{0}\left[-\kappa + \frac{2{N}^{2}}{7} + \kappa^{\prime} + \left(g_{y} - g_{s}\right) - \frac{3}{14} \left(g_{x} - g_{s}\right)\right],\\
A_{z}&=&P_{0}\left[-\kappa - \frac{4{N}^{2}}{7} + \left(g_{z} - g_{s}\right) +
\frac{3}{14} \left(g_{x} + g_{y} - 2g_{s}\right)\right].\nonumber
\end{eqnarray}

Here, ${{g}}_{{s}}$ ($\approx 2.0023$) is the
spin-only value. ${k}$ ($\approx {N}$) is the orbital
reduction factor. $P_{0}$ ($\approx 172 \times 10 ^{-4}$~cm$^{-1}$
\cite{23,24}) is the dipolar hyperfine interaction parameter. The
spin-orbit coupling coefficient for CAS:Cu$^{2+}$
is acquired as the free-ion value $\zeta_{0}$ ($\approx 829$~cm$^{-1}$ \cite{25}) multiplying ${N}$. $\kappa$ and
$\kappa^{\prime}$ are the isotropic and anisotropic core
polarization constants, respectively. The denominators $E_{i}$
(${i} = 1-4$) can be obtained from the energy matrices for a
$3d^{9}$ ion under rhombic symmetry in terms of the cubic
field parameter ${D}_{q}$ and the rhombic field parameters
${D}_{s}$, ${D}_{t}$, ${D}_{\xi}$ and
${D}_{\eta}$:
\begin{eqnarray}
E_{1}&=& 4D_{s}+5D_{t}\,,\nonumber\\
E_{2}&=& 10D_{q}\,,\nonumber\\
E_{3}&=& 10D_{q}+3D_{s}-5D_{t}-3D_{\xi}+4D_{\eta}\,,\\
E_{4}&=& 10D_{q}+3D_{s}-5D_{t}+3D_{\xi}-4D_{\eta}\,.\nonumber
\end{eqnarray}

According to optical spectral studies for Cu$^{2+}$ in
oxides with similar $[\text{Cu}\text{O}_{6}]^{10-}$ cluster
\cite{26,27}, the cubic field parameter $D_{q}$ ($\approx 1050$~cm$^{-1}$) and the orbital reduction factor ${k}$
($\approx 0.84$) can be obtained. From the core polarization constant $\kappa$
($\approx 0.2 - 0.6$ \cite{8,9,20,24}) for Cu$^{2+}$ in
many crystals with similar $[\text{Cu}\text{O}_{6}]^{10-}$ clusters,
one can estimate $\kappa$ ($\approx 0.34$) for the system studied
here. In view of the anisotropic contributions to hyperfine
structure constants from Cu$^{2+}$ $3d$--$3s$ ($4s$) orbital
admixtures, the anisotropic core polarization constants are taken as
$\kappa^{\prime}$ ($\approx 0.021$).

Then, for the impurity Cu$^{2+}$ in CAS single
crystal, we take the Cu$^{2+}$--O$^{2-}$ bond
lengths:
\begin{equation}
{{R}}_{x}\approx 2.05~\text{\AA}, \qquad
{{R}}_{y}\approx 1.91~\text{\AA} \qquad \text{and} \qquad
{{R}}_{x}\approx 2.32~\text{\AA}.
\end{equation}

Substituting the above values into the matrix formulae in
equation (\ref{eq1}), and using the ground state wave function as
follows:
\begin{eqnarray}
\Phi= 0.85\left[0.995|d_{x^{2}-y^{2}}+0.0999|d_{3z^{2}-r^{2}}\right]
\end{eqnarray}
the calculated results (Cal$^{b}$.) of EPR parameters are
shown in table~\ref{tab1}. For comparison, the results (Cal$^{a}$.) based on
the omission of admixture $^{2}{A}_{1g}(\theta)$ and
$^{2}{A}_{1g}(\epsilon)$ states (i.e., taking $\alpha=1$,
$\beta= 0$) are also listed in table~\ref{tab1}.

\begin{table}[!h]
\begin{center}
\caption{The calculated and experimental anisotropic $g$
factors and hyperfine structure constants (in $10^{-4}$~cm$^{-1}$) for Cu$^{2+}$ in
Cd$_{2}$(NH$_{4}$)$_{2}$(SO$_{4}$)$_{3}$ single crystal.\label{tab1}}
\vspace{2ex}
\begin{tabular}{|c| c|c|c|c|c|c|c|}
   \hline\hline
                &$g_{x}$   &$g_{y}$     &$g_{z}$    &$A_{x}$     &$A_{y}$      &$A_{z}$     \\    \cline{1-7}
  Cal$^{a}$.    &$2.078$   & $2.047$    &$2.431$    &$-33.5$       &$-32.0$      &$-103.8$       \\     \cline{1-7}
  Cal$^{b}$.    &$2.145$   & $2.096$    &$2.412$    &$-19.7$       &$-13.1$      &$-101.9$        \\     \cline{1-7}
 Expt.\cite{9}  &$2.144$   & $2.094$    &$2.415$    &$18.7$        &$15.9$       &$97.1$           \\     \cline{1-7}
 \hline\hline
\end{tabular}
\begin{minipage}{0.9\textwidth}
\vspace{2mm}
{\small $^a$ \!Calculations based on the perturbation formulae
but with the omission of admixture $^{2}{A}_{1g}$$(\theta)$ and
\phantom{\small $^{a}$ }$^{2}\!{A}_{1g}$$(\epsilon)$ states (i.e., taking $\alpha$ = 1,
$\beta$ = 0).}\\
{\small $^{b}$ Calculations based on the perturbation formulae
and considering the admixture of $^{2}{A}_{1g}$$(\theta)$
and $^{2}{A}_{1g}$$(\epsilon)$ \phantom{\small $^{b}$} states.}
\end{minipage}
\end{center}
\end{table}

\section{Discussion}
From table~\ref{tab1}, it can be seen that the calculated EPR
parameters are in good agreement with the experimental data. Thus,
the obtained Cu$^{2+}$--O$^{2-}$ bond lengths and
admixture coefficients of $d$-orbitals for the impurity
Cu$^{2+}$ ion in CAS single crystal are reasonable.

1. From table~\ref{tab1}, one can find that the calculated EPR
parameters by using the perturbation formulae and considering
the admixture of the ground states $^{2}{A}_{1g}(\theta)$ and
$^{2}{A}_{1g}(\epsilon)$ are in good agreement with the
experimental data and the results are better than those obtained using the above
formulas while neglecting the admixture of the $d$-orbitals (i.e., $\alpha= 1$, $\beta= 0$). Moreover, it is noted that the mixing
coefficient $\alpha$ ($\approx 0.980$) is very close to that
($\alpha \approx 0.977 - 0.996$ \cite{8,14,22}) based on the
analysis of EPR parameters for similar rhombic
Cu$^{2+}$ centers in many crystals.

2. From table~\ref{tab1}, one can see that the absolute values of the calculated
hyperfine structure constants ${{A}}_{{i}}$ are in
good agreement with the experimental findings, but the signs of all of them
are negative. Actually, the signs of hyperfine structure constants
are very difficult to ascertain. Thus, many experiments give them as
absolute ones \cite{8,24,28,29}. However, negative signs of
constants ${{A}}_{{i}}$ for ${3d}^{n}$ ions in
many crystals were proposed \cite{20,21,22,23}. Thus, the negative
signs of hyperfine structure constants obtained in this work can be
regarded as reasonable.

\section{Conclusion}

The EPR parameters and the local structures of
rhombic Cu$^{2+}$ center in the CAS crystal are
theoretically investigated from perturbation formulae for a
${3d}^{9}$ ion in rhombically elongated octahedra. The
theoretical results based on the above perturbation formulae and
considering the admixture of the ground states
$^{2}{A}_{1g}(\theta)$ and $^{2}{A}_{1g}(\epsilon)$
are in good agreement with the experimental data. The ligand
octahedra around Cu$^{2+}$ are found to suffer a relative
elongation along the C$_{4}$ axis due to Jahn-Teller effect,
which may entirely depress the original compressed
$\text{Cd}(\text{H}_{2}\text{O})_{6}^{2+}$ octahedra in the host
crystals.

\section*{Acknowledgements}

This work was financially supported by Chinese Natural Science
Foundation (grants 11365017,  \linebreak 11465015).

\ukrainianpart

\title{Теоретичні дослідження локальних структур та параметрів електронного парамагнітного резонансу для центру Cu$^{2+}$  \\ в монокристалі Cd$_{2}$(NH$_{4}$)$_{2}$(SO$_{4}$)$_{3}$}

\author{Ч.-І. Лі, Л.Б. Чен, Дж.-Дж. Мао, Кс.-M. Женг}
\address{Школа фізики і електронної інформації, Нормальний університет Шанграо, Шанграо Цзянсі, КНР}

\makeukrtitle

\begin{abstract}
Параметри електронного парамагнітного резонансу (ЕПР)
($g$-фактори $g_{i}$ і константи надтонкої структури ${{A}}_{{i}}$, ${i} =
{x}, {y}, {z}$) інтерпретуються за допомогою формули збурень для іона $3{d}^{9}$
у ромбічно
({D}$_\textrm{2h}$) видовженому восьмиграннику. У формулах  параметри кристалічного поля встановлюються згідно суперпозиційної моделі, де враховується внесок у параметри  ЕПР  від домішування
$d$-орбіталей в основному стані хвильової функції
Cu$^{2+}$. На основі обчислень отримано локальні структурні параметри домішки
центру Cu$^{2+}$  у кристалі
Cd$_{2}$(NH$_{4}$)$_{2}$(SO$_{4}$)$_{3}$ (CAS)
(тобто, ${R}_{{x}}\approx 2.05$~{\AA}, ${R}_{{y}} \approx 1.91$~{\AA},
${R}_{{z}} \approx 2.32$~{\AA}). Теоретичні параметри ЕПР, що базуються на вищезгаданих довжинах
Cu$^{2+}$--$\text{O}^{2-}$ зв'язку у CAS кристалі, демонструють хороше узгодження  з спостережуваними значеннями.
Результати досліджень обговорюються.
\keywords дефектна структура, електронний парамагнітний резонанс, кристал кадмій амонієвого сульфату, легування Cu$^{2+}$
\end{abstract}

\end{document}